\documentclass[10pt,a4paper]{article}

\usepackage{latexsym}
\usepackage{amssymb}

\newif\ifpdf
\ifx\pdfoutput\undefined
   \pdffalse
\else
   \pdfoutput=1
   \pdftrue
\fi

\ifpdf
   \usepackage[pdftex,
               pdftitle={Risk Aversion and Coherent Risk Measures: a Spectral Representation Theorem},
               pdfsubject={Subjective Risk Aversion and Coherent Risk Measures:  a Spectral Representation Theorem},
               pdfkeywords={Expected Shortfall; Risk measure; worst conditional expectation; tail conditional expectation;
               value-at-risk (VaR); conditional value-at-risk (CVaR); tail mean; coherence; quantile;
               sub-additivity},
               pdfauthor={},
               pdfstartview=FitBH,
               pdfview=FitBH,
               colorlinks=true]{hyperref}
\else
\fi

\newtheorem{theorem}{Theorem}[section]

\newtheorem{remark}[theorem]{Remark}

\newtheorem{proposition}[theorem]{Proposition}
\newtheorem{definition}[theorem]{Definition}

\title{Risk Aversion and Coherent Risk Measures: \\ a Spectral Representation Theorem}
\author{Carlo Acerbi \\ \it \small Abaxbank, Corso Monforte 34, 20122 Milano (Italy) }

\date{July 10, 2001}

\begin{document}
\maketitle
\begin{abstract}
We study a  space of coherent risk measures  $M_\phi$ obtained as certain expansions of coherent elementary basis measures. In this space, the concept of  ``Risk Aversion Function'' $\phi$ naturally arises as the spectral representation of each risk measure in a space of functions of confidence level probabilities. We give necessary and sufficient conditions on  $\phi$ for $M_\phi$ to be a coherent measure. We find in this way a simple interpretation of the concept of coherence and a way to map any rational investor's subjective risk aversion onto a coherent measure and vice--versa. We also provide for these measures their discrete versions $M_\phi^{(N)}$ acting on finite sets of $N$ independent realizations of a r.v. which are not only shown to be coherent measures for any fixed $N$, but also consistent estimators of $M_\phi$ for large $N$. Finally, we find in our results  some interesting and not yet fully investigated relationships with certain results known in insurance mathematical literature.

\end{abstract}

\section{Introduction}
 
It was recently discovered \cite{ANS,AT01,AT2,RU01} that $\alpha$--Expected Shortfall $ES_{(\alpha)}$, correctly defined as the {\em ``average of the $100\alpha\%$ worst losses''} of a portfolio, defines, for any chosen confidence level $\alpha\in(0,1]$, a coherent risk measure\footnote{In this paper we will use {\em ``coherent measure''} and {\em ``risk measure''}  as synonymous for the  reasons already explained in ref. \cite{AT01}.}. It is  then natural to wonder whether there exist other {\em ``probability weighted averages''} of the left tail of a distribution which satisfy the axioms of coherency \cite{ADEH97,ADEH99}. In other words we suspect that the $\alpha$--Expected Shortfall might actually be only one possible choice out of a large space of risk measures.

Given some known risk measures it is easy to generate a new risk measure. In fact, it is elementary to prove that a convex combination of risk measures is coherent as well. So, our strategy will be to study the properties of the space of coherent measures generated by the most general convex combination of $\alpha$--Expected Shortfalls. Then we will try to face two distinct questions. The first is whether this space of measures  is in some sense complete, or if there exists, within the framework we are investigating, some risk measure which does not belong to it. The second question is whether the Expected Shortfall plays any special role in this space as a natural choice, or if any measure of this space could equally be a perfectly admissible and legitimate risk measure.

The basic assumption we are making is that some (but essentially any) probability space $(\Omega,\Sigma,\mathbb{P})$ has been chosen for the profit--loss random variables $X_i$ of a set of portfolios $\pi_i$. We will then restrict our analysis to measures of risk which depend on the probability measure $\mathbb{P}$ alone on which however we will not make any restrictive assumption.
 It is important to keep in mind that our investigation will not aim to include all possible coherent measures. Suppose in fact to have  two distinct probability measures $\mathbb{P}_1$ and $\mathbb{P}_2$ for a random variable $X$. Then, is is easy to see that the statistic 
\begin{equation}
\rho(X) = -\max \{E_{\mathbb{P}_1}[X],E_{\mathbb{P}_2}[X]\}
\end{equation}
defines a coherent measure \cite{ADEH97}. The space of risk measures we are going to explore will certainly not contain examples of this sort.

\section{Generating a new class of Risk Measures}

Expected Shortfall  can be used as a basic object for obtaining new risk measures. It is in fact natural to think of the one--parameter family $ES_{(\alpha)}$ (where $\alpha \in (0,1]$ is the confidence level) as a basis for expansions which define a larger class of risk measures.

Remember that a  risk measure is defined by the following ``coherency axioms'' \cite{ADEH97,ADEH99}:
\begin{definition}[Risk Measure]
\label{def:rm}
Consider a set $V$ of real-valued random variables. 
A function $\rho: V \to \mathbb{R}$ is called a risk measure if it is
\begin{enumerate} 
\item monotonous: $X,Y \in V,\ Y \ge X \quad \Rightarrow \quad \rho(Y) \le \rho(X)$,
\item sub--additive: $X, Y, X+Y \in V\quad \Rightarrow \quad \rho(X+Y) \le
\rho(X) + \rho(Y)$,
\item positively homogeneous: $X \in V,\; h > 0, \;h\,X\in V \quad \Rightarrow \quad \rho(h\,X)
= h\,\rho(X)$, 
\item translation invariant: $X \in V, \ a \in \mathbb{R} \quad \Rightarrow \quad \rho(X + a) =
\rho(X) - a$.
\end{enumerate}  
\end{definition}
It is easy to show that an equivalent set of axioms can be obtained by replacing  the monotonicity axiom with the following positivity axiom
{\em
\begin{enumerate} 
\item[{\it ($i^\prime$)}] positive: $X \in V, \ X \geq 0  \quad \Rightarrow \quad \rho(X)\leq 0$   
\end{enumerate} 
}
Our starting point for constructing an expansion of risk measures is the following elementary
\begin{proposition}
\label{prop:convexcombination} 
Let $\rho_i$  be risk measures for $i=1\ldots n$. Then, any convex combination $\rho=\sum_i \alpha_i\,\rho_i$ ($\alpha_i \in\mathbb{R}^+$ and $\sum_i\alpha_i =1$) is a risk measure. Similarly, if $\rho_\alpha$ is  a one--parameter family of risk measures $\alpha\in[a,b]$, then, for any measure $d\mu(\alpha)$ in $[a,b]$ with $\int_a^b  d\mu(\alpha)=1$, the statistic defined as $\rho =\int_a^b  d\mu(\alpha) \rho_\alpha$ is a risk measure.
\end{proposition}
{\bf Proof:} the check is elementary. One uses the requirement $\alpha_i>0$ (or $d\mu(\alpha)>0$)  to check axioms $(i)$ and $(ii)$ and the requirement $\sum_i \alpha_i =1$ (or $\int_a^b  d\mu(\alpha)=1$) to check  axiom $(iv)$.
\hfill $\clubsuit$

Let's now recall the  definition of Expected Shortfall. 
Let $F_X(x)=P[X\leq x]$ be the distribution function of  the profit--loss $X$ of a given portfolio $\pi$ and define  the usual generalized inverse of $F_X(x)$ as\footnote{Actually, any other sensible definition for the inverse of $F_X$ would not alter definition (\ref{eq:es}) and its properties.}
\begin{equation}
F_X^\leftarrow(p) = \inf \{x| F_X(x) \geq p \}
  \label{eq:inv}
\end{equation}
The $\alpha$--Expected Shortfall defined as
\begin{equation}
ES_{(\alpha)}(X) =  -\frac{1}{\alpha}\,\int_0^\alpha F_X^\leftarrow(p) \,dp
  \label{eq:es}
\end{equation}
can be shown \cite{ANS,AT01} to be a risk measure satisfying the axioms of Definition \ref{def:rm}. 

It's worth mentioning that the Expected Shortfall is closely related but not coincident to the notion of Conditional Value at Risk $CVaR_{(\alpha)}$ or Tail Conditional Expectation $TCE_{(\alpha)}$ defined as \cite{ADEH97,ADEH99,Pf00}
\begin{equation}
CVaR_{(\alpha)}(X) = TCE_{(\alpha)}(X)= -E[X| X\leq F_X^\leftarrow(\alpha)]
  \label{eq:cvar}
\end{equation}
In fact, Conditional Value at Risk is not a coherent measure in general. It coincides with  $ES_{(\alpha)}$ (and it is therefore coherent) only under suitable conditions such as the continuity of the probability distribution function $F_X(x)$ (see \cite{AT01} and references therein).

The mathematical tractability of eq. (\ref{eq:es}) suggests  to exploit  Proposition \ref{prop:convexcombination} using $ES_{(\alpha)}$ as the basic building block for defining new coherent measures. Introducing a  measure $d\mu(\alpha)$ on $\alpha \in (0,1]$, and under suitable  integrability conditions, Proposition \ref{prop:convexcombination} ensures that the statistic
\begin{equation} \label{eq:Mmu}
M_{\mu}(X) =  \int_0^1 d\mu(\alpha)\,\alpha\, ES_{(\alpha)}(X) = -\int_0^1 d\mu(\alpha) \int_0^\alpha dp \, F_X^\leftarrow(p)
\end{equation}
is a risk measure as long as the normalization condition 
\begin{equation}\label{normMu}
\int_0^1 \alpha \, d\mu(\alpha)=1
\end{equation} 
is satisfied. Interchanging the integrals thanks to the Fubini--Tonelli theorem 
\begin{equation} \label{eq:Mphi}
M_{\mu}(X) = - \int_0^1 dp F_X^\leftarrow(p)  \int_p^1 d\mu(\alpha) \equiv  - \int_0^1 dp F_X^\leftarrow(p)  \phi(p) 
\equiv M_\phi(X)
\end{equation}
it is easy to see that the parametrization in terms of any measure $d\mu(\alpha)$ can be traded with a parametrization in terms of a decreasing positive {\em ``risk spectrum''} $\phi(p)=\int_p^1 \,d\mu(\alpha)$. The normalization condition eq. (\ref{normMu}) translates into the following normalization condition  for $\phi$
\begin{equation}\label{normphi}
\int_0^1 \phi(p) dp = \int_0^1 dp \int_p^1 \,d\mu(\alpha)=\int_0^1 \,d\mu(\alpha)\, \int_0^\alpha dp =
\int_0^1 \,d\mu(\alpha)\, \alpha  = 1
\end{equation}

In other words, for any measure $d\mu(\alpha)$ satisfying normalization (\ref{normMu}), we have a different risk measure defined by eq. (\ref{eq:Mmu}) which can  also be expressed by eq. (\ref{eq:Mphi}) with $\phi(p)=\int_p^1 \,d\mu(\alpha)$. Conversely, for any decreasing positive function $\phi(p):(0,1]\to\mathbb{R}^+$ satisfying normalization (\ref{normphi}), eq. (\ref{eq:Mphi}) provides a risk measure which can also be expressed by eq. (\ref{eq:Mmu}) with $d\mu(\alpha) = -d\phi(\alpha)$.

Taking a closer look to eq. (\ref{eq:Mphi}) we see that more than a pointwise characterization of $\phi$, we need to define its properties as an element of the normed space ${\cal L}^1([0,1])$ where every element is represented  by a class of functions which differ at most on a subset of $[0,1]$ of zero measure. The norm in this space is given by
\begin{equation}\label{eq:norm}
\Vert \phi\Vert = \int_0^1 \,\vert \phi(p)\vert \, dp
\end{equation}
Different representative functions $\phi_1,\phi_2$ ($\Vert \phi_1 -\phi_2\Vert=0$) of the same element  $\phi \in {\cal L}^1([0,1])$ will in fact define the same measure $M_\phi$.

The properties of monotonicity and positivity of an element of  ${\cal L}^1([0,1])$ cannot be defined pointwise as for functions. Hence, we adopt the following

\begin{definition}\label{def:ni}
We will say that an element $\phi\in{\cal L}^1([a,b])$ is ``positive'' if $\forall I\subset [a,b]$
\begin{equation}\label{eq:po}
\int_I \, \phi(p)\, dp \geq 0
\end{equation}
 We will say that an element $\phi\in{\cal L}^1([a,b])$ is ``decreasing''  if $\forall q\in(a,b)$ and $\forall\epsilon>0$ such that $[q-\epsilon,q+\epsilon]\subset [a,b]$
\begin{equation}\label{eq:ni}
\int_{q-\epsilon}^q \, \phi(p)\, dp \geq \int^{q+\epsilon}_q \, \phi(p)\, dp
\end{equation}
\end{definition}

It is now convenient to give also the following 
\begin{definition}\label{def:ars}
An element $\phi \in {\cal L}^1([0,1])$ is said to be an ``admissible'' risk spectrum if 
\begin{enumerate}
\item[1)] $\phi$ is positive
\item[2)] $\phi$ is decreasing
\item[3)] $\Vert\phi\Vert=1$ 
\end{enumerate}
\end{definition}

From the above discussion we can therefore easily prove the following

\begin{theorem}
\label{prop:sufficiency} 
Let $M_\phi(X)$ be  defined by 
\begin{equation} \label{eq:Mphi2}
M_\phi(X) =  - \int_0^1  F_X^\leftarrow(p) \, \phi(p) \, dp
\end{equation}
with $\phi \in {\cal L}^1([0,1])$.   If $\phi$ is an admissible risk spectrum then $M_\phi(X)$ is a risk measure. 
\end{theorem}

{\bf Proof:} For all admissible risk spectra $\phi \in {\cal L}^1([0,1])$ it is always possible to find a representative positive and decreasing function $\phi(p)$ which  defines a measure $\mu$ on $[0,1]$ by $d\mu(\alpha)= -d\phi(\alpha
) \,$. Then, the coherency of $M_\phi$ follows from  eqs. (\ref{eq:Mmu}) and (\ref{eq:Mphi}) and Prop. \ref{prop:convexcombination}.
 \hfill $\clubsuit$

\begin{remark} \rm 
The integrability conditions of eq. (\ref{eq:Mphi2}) define the space $V_\phi$ of random variables on which  $M_\phi$ is a risk measure. 
\begin{equation}
V_\phi = \{ X | \, \phi\, F_X^\leftarrow \in {\cal L}^1([0,1])\}
\end{equation}
However, in a real--world risk management application  the integral of (\ref{eq:Mphi2}) will always be well defined and finite. For instance, for $M_\phi$ to be finite, it is sufficient to impose that the expectations $E[X^+]=E[\max(X,0)]$ and $E[X^-]=-E[\min(X,0)]$ are finite and that $\phi(p)$ is bounded.
\end{remark}

\section{The Risk--Aversion  Function}

To  understand  the meaning of the function $\phi(p)$ in eq. (\ref{eq:Mphi2}), let's analyze its role in the case of the Expected Shortfall. It is in fact  easy to see that $ES_{(\alpha)}(X)$ can be identified as\footnote{We make use of the Dirac delta function $\delta (x)$ defined by $\int_a^b f(x)\delta(x-c) dx = f(c)$ $\forall c\in (a,b)$.} 
\begin{equation} \label{exp:ES}
ES_{(\alpha)}(X) = M_\mu(X) = M_\phi(X) \hspace{0.5cm} \mbox{with} \hspace{0.5cm} 
\left\{
\begin{array}{rcl}
d\mu(\beta)&=& \frac{1}{\alpha}\, \delta(\alpha-\beta)  \, d\beta\\
\phi(p) &=& \frac{1}{\alpha}\, \mathrm{1}_{\{0\leq p\leq \alpha\}} = \frac{1}{\alpha}\,\theta(\alpha-p)\\
\end{array}
\right.
\end{equation}
Remember that the measure $ES_{(\alpha)}$ represents the {\em ``average of the $100\alpha\%$ worst losses''} of $X$. In other words, this measure averages the possible outcomes contained in the $\alpha$--left tail of the r.v. $X$ with equal weights. Looking at eq. (\ref{exp:ES}), one realizes that the $\phi$  function is nothing but the weighting function in this average which in this case is simply uniform in $p\in (0,\alpha]$ and zero elsewhere. In the general case, the function $\phi(p)$ in eq. (\ref{eq:Mphi2})  assigns in fact different weights $\phi(p)$ to  different ``$p$--confidence level slices'' of the left tail. Normalization $\Vert\phi\Vert=1$ in turn ensures that the weights in the average sum up to $1$.

The fact that an admissible risk spectrum $\phi(p)$ is  a decreasing monotonic function in $p$ provides us with an  intuitive insight of the concept of coherence. In fact, Theorem \ref{prop:sufficiency} simply teaches us  the following reasonable rule:  {\em ``a measure is coherent if it assigns bigger weights to worse cases''}. 

Any rational investor can express her subjective risk aversion in drawing a different profile for the weighting function $\phi$. To attain coherency she has just to restrict the choice of this function to be positive, decreasing and normailzed to one on the interval $(0,1]$. Within these constraints, however, any choice for $\phi$ will represent a perfectly legitimate attitude toward risk. The choice of $\alpha$--Expected Shortfall, for instance, could  not be  satisfactory for any $\alpha$ to a certain  investor who wants to distinguish portfolios which might differ even just at a low risk confidence level. For such an investor, a non--vanishing $\phi(p)$ function on all the confidence level domain $p\in(0,1]$ would be more appropriate.

In general, in the space of measures spanned by all possible admissible risk spectra via Theorem \ref{prop:sufficiency}, no natural choice is provided by purely financial arguments and 
the function $\phi$ appears as the instrument by which an investor can express her subjective attitude toward risk.
We will therefore give the following 

\begin{definition}[Risk Aversion Function and Spectral Risk Measure]
\label{def:RAF}  An admissible risk spectrum $\phi \in {\cal L}^1([0,1])$ will be called the ``Risk Aversion Function'' of the risk measure 
\begin{equation} 
M_\phi(X) \equiv  - \int_0^1  F_X^\leftarrow(p) \, \phi(p) \, dp
\end{equation}
The risk measure $M_\phi$, in turn  will be called the ``spectral risk measure'' generated by $\phi$.
\end{definition}

The following question now arises: is the admissibility of $\phi$ also necessary for coherency?  As a significative example let's consider  the case of Value at Risk:

It is not difficult to see that $\mathrm{VaR}_{(\alpha)}(X)= - F_X^\leftarrow(\alpha)$ can also be expressed as\footnote{We use the usual first derivative $\delta^\prime(x)$ of a Dirac delta. It may be thought of as a formal object on which we can integrate by parts to get rid of the derivative on $\delta$. So $\int_a^b f(x)\delta^\prime(x-c) dx =-\int_a^b f^\prime(x)\delta(x-c) dx = -f^\prime(c)$ $\forall c\in (a,b)$} 
\begin{equation} \label{exp:VaR}
\mathrm{VaR}_{(\alpha)}(X) = M_\mu(X) = M_\phi(X) \hspace{.5cm} \mbox{with} \hspace{.5cm} 
\left\{
\begin{array}{rcl}
d\mu(\beta)&=& - \delta^\prime (\beta-\alpha)  \, d\beta\\
\phi(p) &=& \delta(p-\alpha) \\
\end{array}
\right.
\end{equation}
For this expression, however, Theorem \ref{prop:sufficiency} is not applicable since $\phi$  is not  a decreasing  function in $p$ and therefore it is not an admissible risk spectrum. Indeed, it is well known that $\mathrm{VaR}_{(\alpha)}$ is not a risk measure, due to its lack of subadditivity.

This graphical interpretation of the non--coherency of VaR shows that, in the class of measures we are exploring, $\mathrm{VaR}_{(\alpha)}$ is maybe the less appropriate one since its $\phi(p)$ is somehow the furthest example one can imagine from the concept of a decreasing function: it is a function which is zero everywhere but in $p=\alpha$ where it blasts to infinity. The pictorial interpretation also illustrates the fact that $\mathrm{VaR}_{(\alpha)}$ actually doesn't take into account at all the losses associated to the tail, focusing only on their threshold value. Its risk aversion function displays the attitude of an incoherent investor who is only concerned about the threshold level of her  worst $100\alpha\%$ losses and neglects at all the losses themselves.

This example enforces our belief that if $\phi$  is not an admissible risk spectrum then $M_\phi$ cannot be a risk measure. In the next chapter we will prove that in fact this is the case.

\section{Necessity of the admissibility of $\phi$}

In the following we will  prove that for the measure $M_\phi$ to be a risk measure the  conditions of admissibility of the risk aversion function $\phi$ are not only sufficient but also necessary. We want in other words to prove the following central result of the paper:

\begin{theorem}
\label{prop:iff} 
Let $M_\phi(X)$ be  defined by 
\begin{equation} \label{eq:Mphi3}
M_\phi(X) =  - \int_0^1  F_X^\leftarrow(p)\,  \phi(p)\, dp 
\end{equation}
with $\phi \in {\cal L}^1([0,1])$.  $M_\phi(X)$ is a risk measure if and only if $\phi$ is an admissible risk spectrum.
\end{theorem}
{\bf Proof:} 

Necessity of condition $1)$ of definition \ref{def:ars}. Suppose that $\exists I= [q_1,q_2] \subset (0,1)$ where
\begin{equation}\label{eq:hypopo}
\int_I \, \phi(p)\, dp < 0
\end{equation}
Consider two random variables $Y>X$ on  a probability space $(\Omega,\Sigma,\mathbb{P})$ with elementary events $\Omega = \{\omega_1,\omega_2,\omega_3\}$ and suppose that the probability $\mathbb{P}$ is defined by
\begin{center}
\begin{tabular}{|c|ccc|}\hline 
\multicolumn{1}{|c}{$\omega$} & \multicolumn{1}{|c}{$\mathbb{P}(\omega)$}	&\multicolumn{1}{c}{$X(\omega)$}	& \multicolumn{1}{c|}{$Y(\omega)$} 	 \\ \hline 
$\omega_1$	&$q_1$ 		&$X_{1}$	&$Y_{1}=X_1$	\\
$\omega_2$	&$q_2-q_1$		&$X_{2}$	&$Y_{2}=X_2+a$		\\
$\omega_3$	&$1-q_2$		&$X_{3}$	&$Y_{3}=X_3$	\\	\hline
\end{tabular}
\end{center}
where we suppose $X_1<X_2<X_3$, $Y_1<Y_2<Y_3$, and $a>0$, so that
\begin{center}
\begin{tabular}{|c|cc|}\hline 
\multicolumn{1}{|c}{$p$} & \multicolumn{1}{|c}{$F^\leftarrow_X(p)$}	&\multicolumn{1}{c|}{$F^\leftarrow_Y(p)$}	  	 \\ \hline 
$p\in(0,q_1]$	 	&$X_{1}$	&$Y_{1}$		\\
$p\in(q_1,q_2]$		&$X_{2}$	&$Y_{2}$		\\
$p\in(q_2,1]$		&$X_{3}$	&$Y_{3}$		\\ \hline
\end{tabular}
\end{center}
Now, it is easy to compute 
\begin{eqnarray}
M_\phi(Y)-M_\phi(X) &=& - \int_0^1 \phi(p) \left(F^\leftarrow_Y(p)-F^\leftarrow_X(p) \right) dp \\
&=& -  a \,\int_{I} \phi(p) dp \nonumber \\
&>& 0   \nonumber 
\end{eqnarray}
This shows that eq. (\ref{eq:hypopo}) contradicts axiom $i)$ in Definition \ref{def:rm}. 

Necessity of condition $2)$ of definition \ref{def:ars}. Suppose that $\exists q\in (0,1)$ and $\epsilon>0$ such that $[q-\epsilon,q+\epsilon]\in(0,1)$ and
\begin{equation}\label{hypo}
\int_{q-\epsilon}^q \, \phi(p)\, dp < \int^{q+\epsilon}_q \, \phi(p)\, dp
\end{equation}
Consider three random variables $X+Y=Z$ defined on  a probability space $(\Omega,\Sigma,\mathbb{P})$ with elementary events $\Omega = \{\omega_1,\omega_2,\omega_3,\omega_4\}$ and suppose that the probability $\mathbb{P}$ is defined by
\begin{center}
\begin{tabular}{|c|cccc|}\hline 
\multicolumn{1}{|c}{$\omega$} & \multicolumn{1}{|c}{$\mathbb{P}(\omega)$}	&\multicolumn{1}{c}{$X(\omega)$}	& \multicolumn{1}{c}{$Y(\omega)$} & \multicolumn{1}{c|}{$Z(\omega)$}	 \\ \hline 
$\omega_1$	&$q-\epsilon$ 	&$X_{1}$	&$Y_{1}$	&$Z_{1} = X_{1}+Y_{1}$	\\
$\omega_2$	&$\epsilon$		&$X_{2}$	&$Y_{3}$	&$Z_{2} = X_{2}+Y_{3}$	\\
$\omega_3$	&$\epsilon$		&$X_{3}$	&$Y_{2}$	&$Z_{3} = X_{3}+Y_{2}$	\\
$\omega_4$	&$1-q-\epsilon$	&$X_{4}$	&$Y_{4}$	&$Z_{4} = X_{4}+Y_{4}$	\\ \hline
\end{tabular}
\end{center}
Subscripts in $X,Y,Z$ have been chosen for ordering the possible outcomes, so  $X_i<X_j$ if $i<j$ and so on. We have deliberately chosen the twist $Y(\omega_2)=Y_3$, $Y(\omega_3)=Y_2$  and we supposed $X_{2}+Y_{3}<X_{3}+Y_{2}$. Now it is easy to compute
\begin{center}
\begin{tabular}{|c|ccc|}\hline 
\multicolumn{1}{|c}{$p$} & \multicolumn{1}{|c}{$F^\leftarrow_X(p)$}	&\multicolumn{1}{c}{$F^\leftarrow_Y(p)$}	& \multicolumn{1}{c|}{$F^\leftarrow_Z(p)$} 	 \\ \hline 
$p\in(0,q-\epsilon]\equiv I_1$	 	&$X_{1}$	&$Y_{1}$	&$Z_{1}$	\\
$p\in(q-\epsilon,q]\equiv I_2$		&$X_{2}$	&$Y_{2}$	&$Z_{2}$	\\
$p\in(q,q+\epsilon]\equiv I_3$		&$X_{3}$	&$Y_{3}$	&$Z_{3}$	\\
$p\in(q+\epsilon,1]\equiv I_4$		&$X_{4}$	&$Y_{4}$	&$Z_{4}$	\\ \hline
\end{tabular}
\end{center}
and
\begin{eqnarray}
M_\phi(Z)-M_\phi(X)-M_\phi(Y) &=& - \int_0^1 \phi(p) \left(F^\leftarrow_Z(p)-F^\leftarrow_X(p)-F^\leftarrow_Y(p) \right) dp
\\
&=& - \sum_{i=1}^4 \, \int_{I_i} \phi(p) \left( Z_i -X_i-Y_i\right) dp \nonumber \\
&=& - (Y_3-Y_2) \left( \int_{I_2} \phi(p) dp -\int_{I_3} \phi(p) dp  \right)\nonumber \\
&>& 0 \nonumber 
\end{eqnarray}
This shows that if eq. (\ref{hypo}) holds, then $M_\phi$ violates axiom $ii)$ in Definition \ref{def:rm}.

Necessity of $\Vert\phi\Vert=1$.   For any  r.v. $X$ and $a\in \mathbb{R}$ we have $F_{X+a}^\leftarrow(p)= F_{X}^\leftarrow(p)+a$. Then
\begin{equation}
M_\phi(X+a) = -\int_0^1\,\phi(p) F_{X+a}^\leftarrow(p)\, dp= M_\phi(X) - a \int_0^1\,\phi(p)\, dp
\end{equation}
which satisfies axiom $(iv)$ in Definition \ref{def:rm} only if $\int_0^1\phi(p) dp = 1$.  \hfill $\clubsuit$

Theorem \ref{prop:iff} provides a one--to--one correspondence between risk aversion functions $\phi \in {\cal L}^1([0,1])$ and spectral risk measures $M_\phi$. All the possible risk measures which can be generated by the expansion (\ref{eq:Mphi3}) are spanned by all the possible admissible risk spectra $\phi$. In this sense, this space of risk measures can be said to be complete.

\begin{remark} \rm As pointed out to me by D. Tasche,  it is interesting to notice that in insurance mathematical literature, there exists a result which is amazingly similar to Theorem \ref{prop:iff}, namely Theorem 4 in reference \cite{Wang}.  It is surprising to notice that this  paper dates back to 1995 and it is therefore older than references \cite{ADEH97,ADEH99} where the notion of coherent measure of risk was introduced in financial mathematics. The similarity and differences between our and Wang's approaches deserves a deeper investigation which will be made in a forthcoming publication \cite{ATfuture}. 

The scope of the present investigation is also enlarged by the results of ref. \cite{BLS00}, where the introduction of Expected Shortfall as a risk measure was motivated by second--order stochastic dominance. This paper, in fact, explores a connection between coherent measures and expected utility theory\footnote{Some care must be taken since in this paper a tacit assumption of continuity of the distribution functions is made, under which the identification of Expected Shortfall as defined in eqs. (3) and (8) is made legitimate.}.
\end{remark}

\section{From theory to practice} \label{discrete}

Despite its appearance, the risk measure $M_\phi$  of Theorem \ref{prop:iff} is in fact a very simple object to be used in practice. The integral of eq.  (\ref{eq:Mphi3}) is however computable only when an explicit analytical expression for the inverse cumulative distribution function $F_X^\leftarrow(p)$ is available. In a real world risk management system this is typically the case only if the approach chosen  for the probability distributions is  parametric.

In fact, the most straightforward method for evaluating $M_\phi$ is  not by its integral definition, but rather by the estimator $M_\phi^{(N)}$ on a sample of $N$ i.i.d. realizations $X_1,\ldots,X_N$ of the portfolio profit--loss $X$. To define it we need to introduce the ordered statistics $X_{i:N}$ given  by the ordered values of the N-tuple $X_1,\ldots,X_N$. In other words: $\{X_{1:N},\ldots,X_{N:N}\}=\{X_1,\ldots,X_N\}$ and  $X_{1:N}\leq X_{2:N}\ldots\leq X_{N:N}$.

\begin{definition} Let $X_1,\ldots,X_N$ be $N$ realizations of a r.v. $X$. For any given $N$--tuple of weights $\phi_{i=1,\ldots,N}\in \mathbb{R}$ we define the statistics
\label{def:Mn}
\begin{equation}
M_\phi^{(N)}(X) = - \sum_{i=1}^N \; X_{i:N}\,\phi_i
\end{equation}
We will call $M_\phi^{(N)}$ the spectral risk measure generated by $\phi_i$. 
\end{definition}

The discrete version of ``admissible risk spectrum''   sounds

\begin{definition}\label{def:discretera} An  $N$--tuple  $\phi_{i=1,\ldots,N}\in \mathbb{R}$ is said to be an ``admissible'' risk spectrum if 
\begin{enumerate}
\item[1)] $\phi_i\geq 0$ ($\phi_i$ is positive)
\item[2)] $\phi_i\geq \phi_j$ if $i<j$ ($\phi_i$ is decreasing)
\item[3)] $\sum_i\,\phi_i =1$ 
\end{enumerate}
\end{definition}

We can now prove the discrete version of  Theorem \ref{prop:iff}

\begin{theorem}
\label{prop:discreteiff} 
The spectral risk measure $M_\phi^{(N)}$  of Definition (\ref{def:Mn}) is a risk measure for any fixed $N\in\mathbb{N}$ if and only if $\phi_i$ is an admissible risk spectrum.
\end{theorem}
{\bf Proof:} This is in fact a special case of Theorem \ref{prop:iff}. To see why, we notice that given $N$ independent realizations $X_{i=1,\ldots,N}$ of a r.v. $X$, the equation $M_\phi^{(N)}(X)=M_\phi(X)$ holds provided that in computing $M_\phi(X)$ we adopt for $X$ the ``empirical'' probability distribution function
\begin{equation}\label{FN}
F_X^{(N)}(x) = \frac{1}{N} \sum_{i=1}^N \mathrm{1}_{\{x\geq X_i\}}
\end{equation}
and the risk spectrum $\phi:(0,1]\to \mathbb{R}^+$
\begin{equation} \label{phiN}
\phi(p) = N \sum_{i=1}^N \phi_i \, \mathrm{1}_{\{Np\in (i-1,i]\}}
\end{equation}
which is admissible if and only if $\phi_{i=1,\ldots,N}$ is admissible. Both sufficiency and necessity therefore follow immediately. 
\hfill $\clubsuit$

Theorem \ref{prop:discreteiff} has a wide range of applicability, since it provides a risk measure for a sample of $N$ realizations of a random variable $X$. The coherency of the measure is not related to some law of large numbers, because the theorem holds for any finite $N\in\mathbb{N}$. This result is immediately applicable in any scenario--based risk management system (parametric Montecarlo scenarios, historical scenario simulation and so on \ldots).

In practice, an investor should choose her own risk averse function $\phi(p)$ to assess her risks independently of the number of scenarios available for the estimation of $M_\phi$. Here, for sake of concretness, we can consider $\phi(p)$ as a positive decreasing normalized  function rather than an abstract element of ${\cal L}^1([0,1])$.
Given $\phi(p)$ and fixed a number $N$ of scenarios, the most natural choice for an admissible sequence $\phi_i$ is given  by
\begin{equation}
\phi_i = \frac{\phi(i/N)}{\sum_{k=1}^N\, \phi(k/N)} \hspace{1cm}i=1,\ldots,N
\end{equation}
This expression in particular satisfies $\sum_i \phi_i = 1$ for any finite $N$. The investor can then use the spectral risk measure  $M_\phi^{(N)}$ generated by this sequence as a risk measure, since Theorem \ref{prop:discreteiff} ensures its coherence for any finite $N$.

However, we can  prove that in fact $M_\phi^{(N)}$ is also a {\em consistent estimator} which converges to $M_\phi$ with probability 1 for $N\to\infty$. To prove this result we need some integrability conditions on $F^\leftarrow_X$ and on $\phi$. For our purposes it will be sufficient to impose  that the expectations $E[X^+]$ and $E[X^-]$ are finite and that the function $\phi(p)$ is bounded. These conditions, from a practical point of view are always satisfied in a risk management application. The theorem can be proved also under weaker conditions (see ref. \cite{Zwet}).

\begin{theorem}
\label{prop:estimator}
Let $X$ be a r.v. with $E[X^+]<\infty$ and $E[X^-]<\infty$ and let $M_\phi$ be the  spectral risk measure generated by some admissible risk spectrum $\phi \in {\cal L}^1([0,1])$ of which $\phi(p)$ is a representative positive decreasing function such that  $\sup_{p\in(0,1]}\phi(p)<\infty$.
Then, if $M_\phi^{(N)}$ is the risk measure generated by the sequence
\begin{equation}
\phi_i = \frac{\phi(i/N)}{\sum_{k=1}^N\, \phi(k/N)}\hspace{5mm} i=1\ldots N
\end{equation}
$M_\phi^{(N)}(X)$ converges  to  $M_\phi(X)$ for $N\to\infty$ with probability 1.
\end{theorem}
{\bf Proof:} This theorem is a special case of Theorem 3.1 of ref. \cite{Zwet} with $t_0=0$, $t_1=1$, $p_1=\infty$, 
$J(t) = \phi(t)$, $g(t)= F_X^\leftarrow(t)$ and $J_N(t) = N\, \phi_i$ for  $(i-1)/N <t\leq i/N $. \hfill $\clubsuit$

We have then shown that  $M_\phi^{(N)}$ provides not only a coherent measure for any fixed $N$, but also a consistent way for estimating, for large number of scenarios the risk measure $M_\phi$. In a scenario--based risk management system this gives the possibility of estimating any spectral risk measure $M_\phi$ in a straightforward and effortless fashion.

\section{Conclusions}

In this paper we have defined a complete space of coherent measures of risk (the ``spectral risk measures'') 
depending on the probablility measure $\mathbb{P}$  and we have provided for each of these measures a spectral representation in terms of its risk aversion function $\phi$ (Theorem \ref{prop:iff}). This representation is not only a constructive recipe for obtaining all the measures of this space, but provides us with an intuitive insight of the concept of coherency. The space of coherent measures $M_\phi$ is in fact in one--to--one correspondence with those elements $\phi \in {\cal L}^1([0,1])$ which are identified as the set of admissible risk spectra $\phi$ (Definition \ref{def:ars}).

We also obtain analogous results for risk measures $M_\phi^{(N)}$ defined as functions of $N$ realizations of a r.v. $X$. We show in fact that for any fixed $N$, these measures are coherent and  the space of these measures is completely spanned by the set of all discrete admissible risk spectra $\phi_i$ (Definition \ref{def:discretera} and Theorem \ref{prop:discreteiff}).

Furthermore, we show that $M_\phi^{(N)}$ is not only a risk measure itself, but also a consistent estimator, for $N\to\infty$ of $M_\phi$ if the  risk spectrum $\phi_{i=1,\ldots,N}$ is chosen as the natural discretization of $\phi$ (Theorem \ref{prop:estimator}).

We have therefore provided a scheme where the subjective risk aversion of an investor can be encoded in a function $\phi(p)$ defined on all the possible confidence levels $p\in [0,1]$. This function in turn generates a spectral risk measure which gives a coherent assessment of risks. 

From a purely financial point of view we do not see any natural choice in the space of admissible risk aversion functions $\phi$, nor any reason to reject any subset of the space of risk measures they span. Every risk measure in this space appears to be a legitimate candidate for a risk measure.

It is on a subjective ground that the choice among the measures of this space has to  be made. The actual shape of the portfolio profit--and--loss distribution and the subjective risk aversion of the investor may help to select out some optimal choice in a specific case.

Any of the measures of risk defined in this paper can be implemented in a risk management system in an elementary way, with  no computational effort, following in particular the approach of Section \ref{discrete}   

Finally, we find an interesting connection between the notion of coherent risk measure used in financial mathematics and  similar concepts introduced long ago in insurance and actuarial mathematics. The connection  is provided by the strict analogy between our main result (Theorem \ref{prop:iff}) and Theorem 4 of Reference \cite{Wang}.

{\bf Acknowledgements:} I want to thank Dirk Tasche who criticized a naive early version of this paper and gave me precious comments and suggestions.


\newpage
\sloppy


\begin{thebibliography}{99}
\bibitem{ANS} \textsc{Acerbi, C., Nordio, C., Sirtori, C. (2001)} Expected Shortfall as a Tool for Financial Risk Management. Working paper.
\ifpdf
\href{http://www.gloriamundi.org/var/wps.html}
{\tt http://www.gloriamundi.org/var/wps.html}
\else
{\tt http://www.gloriamundi.org/var/wps.html}
\fi
\bibitem{AT01} \textsc{Acerbi, C., Tasche, D. (2001)} On the Coherence of Expected Shortfall. 
Working paper.
\ifpdf
\href{http://www.gloriamundi.org/var/wps.html}
{\tt http://www.gloriamundi.org/var/wps.html}
\else
{\tt http://www.gloriamundi.org/var/wps.html}
\fi
\bibitem{AT2} \textsc{Acerbi, C., Tasche, D. (2001)} Expected Shortfall: a natural coherent alternative to Value at Risk. 
Working paper.
\ifpdf
\href{http://www.gloriamundi.org/var/wps.html}
{\tt http://www.gloriamundi.org/var/wps.html}
\else
{\tt http://www.gloriamundi.org/var/wps.html}
\fi
\bibitem{ATfuture} \textsc{Acerbi, C., Tasche, D. } {\em Work in progress.} 
%
\bibitem{ADEH97} \textsc{Artzner, P., Delbaen, F., Eber, J.-M., Heath, D. (1997)} Thinking
coherently. RISK \textbf{10} (11).
%
\bibitem{ADEH99} \textsc{Artzner, P., Delbaen, F., Eber, J.-M., Heath, D. (1999)} 
Coherent measures of risk. Math. Fin. \textbf{9}(3), 203--228.

\bibitem{BLS00} \textsc{Bertsimas, D., Lauprete, G.-J., Samarov, A. (2000)} 
Shortfall as a risk measure: properties, optimization and applications. 
Working paper, Sloan School of Management,
MIT, Cambridge.


\bibitem{Pf00} \textsc{Pflug, G. (2000)} Some remarks on the value-at-risk
and the conditional value-at-risk. In, Uryasev, S.
(Editor). 2000. Probabilistic Constrained Optimization: Methodology and Applications. Kluwer Academic
Publishers.
\ifpdf
\href{http://www.gloriamundi.org/var/pub.html}
{\tt http://www.gloriamundi.org/var/pub.html}
\else 
\texttt{http://www.gloriamundi.org/var/pub.html}
\fi
\bibitem{RU01} \textsc{Rockafellar, R.T., Uryasev, S. (2001)} Conditional Value-at-Risk for general loss distributions. Research report 2001-5, ISE Depart., University of Florida.
\ifpdf
\href{http://www.ise.ufl.edu/uryasev/} 
{\tt http://www.ise.ufl.edu/uryasev/}
\else
\texttt{http://www.ise.ufl.edu/uryasev/}
\fi
\bibitem{Ur00} \textsc{Uryasev, S. (2000)} \emph{Conditional Value-at-Risk: Optimization Algorithms and Applications.}
Financial Engineering News \textbf{2} (3).\\
\ifpdf
\href{http://www.gloriamundi.org/var/pub.html}
{\tt http://www.gloriamundi.org/var/pub.html}
\else 
\texttt{http://www.gloriamundi.org/var/pub.html}
\fi
\bibitem{Zwet} \textsc{van Zwet, W. R. (1980)} A strong law for linear functions of order statistics, Ann. Probab. 
 \textbf{8}, 5 986--990.
\bibitem{Wang} \textsc{Wang, S. (1996)} Premium Calculation by Transforming the Layer Premium Density
Astin Bulletin \textbf{26}, 71--92.
\end{thebibliography}
\end{document}